\documentclass[letterpaper, twocolumn,superscriptaddress, showpacs,preprintnumbers,amsmath,amssymb] {revtex4}
\usepackage{graphicx}
\usepackage{dcolumn}
\usepackage{bm}
\usepackage{hyperref}
\hypersetup{
    colorlinks,
    citecolor=red,
    filecolor=blue,
    linkcolor=blue,
    urlcolor=blue
}
\usepackage{amsbsy}
\newcommand{\be}{\begin{equation}}
\newcommand{\ee}{\end{equation}}
\newcommand{\bsig}{\boldsymbol{\sigma}}
\newcommand{\btau}{\boldsymbol{\tau}}

\usepackage{cancel}
\begin{document}

\title{A phenomenology of certain many-body-localized systems}

\author{David A. Huse}
\affiliation{Physics Department, Princeton University, Princeton, NJ 08544, USA}
\author{Vadim Oganesyan}
\affiliation{Department of Engineering Science and Physics, College of Staten Island, CUNY, Staten Island, NY 10314, USA}
\begin{abstract}
We consider isolated quantum systems with all of their many-body eigenstates
%Anderson
localized.
We define a sense in which such systems are integrable,
%We argue that such systems are in a certain sense integrable,
and discuss a method
%to in principle \textcolor{red}{(to remove if put in practice...)}
for finding
their localized conserved quantum numbers (``constants of motion'').  These localized operators are interacting pseudospins and
%that
are subject to
dephasing but not to dissipation, so any quantum states of these pseudospins can in principle be recovered
via (spin) echo procedures.  We also discuss the spreading of entanglement in many-body localized systems,
which is another aspect of the dephasing due to interactions between these localized conserved operators.
%Although the many-body eigenstates of such Hamiltonians generally have only area-law entanglement,
%interactions and dephasing of an initial pure product state cause the entanglement to ``propagate'', spreading
%logarithmically with time.
\end{abstract}
\date{\today}
%\pacs{67.85.Lm, 34.50.-s, 67.80.dj}
\maketitle

Isolated quantum many-body systems with short-range interactions and static randomness may be
in a many-body localized phase where they do not thermally equilibrate under their own dynamics.
While this possibility was pointed out long ago by Anderson \cite{pwa}, such localization of highly-excited states
in systems with interactions did not receive a lot of attention %\footnote{dykman -- "strong many-body localization"}
until after Basko, {\it et al.} \cite{baa} forcefully
brought the subject into focus.  Isolated systems in the many-body localized phase have strictly
zero thermal conductivity \cite{baa}, so if some energy is added to the system locally, it only excites localized
degrees of freedom and does not diffuse, even when the system's energy density corresponds to a nonzero
(even infinite \cite{oh}) temperature.

We expect that the many-body eigenstates of a system's Hamiltonian in the localized phase
are product states of localized degrees of freedom, with some short-range ``area-law''
entanglement between the ``bare'' local degrees of freedom.  %One might naively think that because there is no transport in this phase and its eigenstates have
%only short-range entanglement there will be only a limited growth of the system's entanglement under its
%unitary time evolution.  However, this is not true, as we discuss below, and as has been indicated \cite{znid,bard} and understood in recent
%work \cite{vosk}.
One goal of this paper is to explore how one can define suitably ``dressed'' localized pseudospin operators in terms of which the
many-body eigenstates within the localized phase are indeed precisely product states with zero entanglement.
When the Hamiltonian is then expressed in terms of
these dressed localized pseudospins it has exponentially decaying long-range interactions, and it is these long-range interactions that cause
decoherence and dephasing of %the
local observables in the many-body insulator. %ized pseudospins.
These interactions also cause the spreading of
entanglement for a %to appear dynamically from an %generic
nonentangled initial product state of the bare spins, as has been seen and explored in Refs. \cite{znid,bard,vosk,iyer,spa}.

To be concrete, assume we have a system of $N$ spin-1/2's $\{\bsig_i\}$ on some lattice (say, in one, two or three dimensions).
For an example, see, e.g., Ref. \cite{pal}.
Our system has a specific random Hamiltonian $H$ that contains only
short-range interactions and strong enough static random fields on each spin so that, %\tred{essentially}
with probability one in the limit of large $N$, all $2^N$ many-body eigenstates of this $H$ are localized.
%\footnote{\tred{Localization of a many-body eigenstate can be defined based on whether said eigenstate contributes to zero frequency transport,
%whether it posesses a discrete local spectrum of excitations, but also if it also has finite bipartite entablement in the thermodynamic limit}}
%in the limit of large $N$ \tred{, i.e. with unit probability}.
The construction we present below should be readily generalizable to local operators with more than two states.
It should also be generalizable to systems where the dominant strong randomness is instead the spin-spin interactions rather than random fields.
In those cases, the pseudospins we will construct may instead be localized domain wall operators \cite{z2} or spin-exchange operators \cite{vosk}
and the lowest-energy mode may be either a global symmetry mode \cite{z2} or bilocalized between distant sites \cite{vosk}.
However, as we discuss below, it %may not be possible to
is not straightforward to
generalize our construction to weaker disorder if the spectrum of $H$ then contains both localized and thermal many-body eigenstates. % remains unclear.

For this specific fully localized
system of $N$ interacting spin-1/2's with Hamiltonian $H$, we will define $N$ suitably-dressed local spin-1/2 pseudospins $\{\btau_i\}$.  We call the bare spins $\{\bsig_i\}$
``p-bits'' (p=physical) and the dressed spins $\{\btau_i\}$ ``l-bits'' (l=localized).  By construction (see below), the full $2^N$-dimensional state space of our system is
an outer product of all of the $N$ independent 2-dimensional state spaces of the l-bits, and the $z$ components of the l-bit Pauli operators $\{\tau^z_i\}$ are the constants of motion of this integrable system.  The $\{\tau^z_i\}$ all commute
with $H$ (and with each other), so each many-body eigenstate of $H$ is one of the $2^N$ simultaneous eigenstates of all of the $\tau^z_i$'s.
Such a construction defining the l-bit operators can always be made \cite{lych}, in fact, there are $(2^N)!$ possible ways to do it,
since there are that many one-to-one assignments between the $2^N$ many-body eigenstates of $H$ and the $2^N$ simultaneous eigenstates of all of the
$\{\tau^z_i\}$'s.  However, almost all such assignments will fail to produce localized l-bits.
Thus we want to choose the ``best'' such definition of the l-bits, which should be the one that has the l-bit operators $\{\btau_i\}$
each most localized near its ``site'' $i$.  We will now attempt to more precisely define a
possible criterion for the ``best'' such definition of the l-bits.

First, let's look at the definitions of the localized operators $\btau_i$ for one specific location $i$.
Each of the many-body eigenstates of $H$ is specified to be a simultaneous eigenstate of all of the
$\{\tau^z_j\}$'s, with one particular one-to-one assignment now assumed.  Of these eigenstates, half have $\tau^z_i=+1$; let's call those states
$\{|\alpha\rangle\}$.  For each of these $2^{(N-1)}$ states $|\alpha\rangle$ we can flip l-bit $i$ to make the state $|\bar\alpha\rangle=\tau^x_i|\alpha\rangle$, which is, by construction, also a many-body eigenstate of $H$ and has $\tau^z_i=-1$, while all the other $\tau^z_j$'s have the same value in $|\alpha\rangle$ and $|\bar\alpha\rangle$.  Thus we can define the l-bit Pauli operators at location $i$ as
\begin{equation}
\tau^z_i=\sum_\alpha (|\alpha\rangle\langle\alpha|-|\bar\alpha\rangle\langle\bar\alpha|)~,
\end{equation}
\begin{equation}
\tau^x_i=\sum_\alpha (|\alpha\rangle\langle\bar\alpha|+|\bar\alpha\rangle\langle\alpha|)~,
\end{equation}
\begin{equation}
\tau^y_i=-i\sum_\alpha (|\alpha\rangle\langle\bar\alpha|-|\bar\alpha\rangle\langle\alpha|)~.
\end{equation}
Note that each $\tau^z_i$ consists of a sum of projectors on to many-body eigenstates of $H$ and thus
commutes with $H$ and with $\tau^z_j$ for all other sites $j$.  To define the l-bit operators at all other locations, just repeat the above.

Next we want to express each such l-bit operator in terms of the bare p-bit operators.  The full set of all linear operators on our $2^N$-dimensional
state space is $4^N$ linearly-independent operators.  One way to list these operators is all $4^N$ composite operators that can made as (outer) products of one
p-bit Pauli operator $\{\sigma^a_i\}$ from each site, where $a=0,x,y$ or $z$, and $0$ denotes the identity operator for that p-bit.  Of these $4^N$ p-bit product operators, only of order $N$ of them are ``local'' operators that consist of the identity operator at every site except at one or a few sites that are all near each other.  The vast majority of the set of all operators are, on the other hand, ``global'' operators that operate nontrivially and simultaneously on of order $N$ of the p-bits.  For a given Hamiltonian $H$, and a given assignment of all its many-body eigenstates to eigenstates of the l-bits, the l-bit operators as defined above can each be expanded in terms of these p-bit product operators.

%Thus for each possible choice of l-bit operators (each one-to-one matching of the many-body eigenstates of $H$ with the l-bit eigenstates) we can expand each l-bit
%operator in terms of these p-bit product operators.

Each p-bit product operator has a range $\ell$, which can be defined as the distance between the two farthest-apart non-identity local p-bit operators that it contains.  Thus we can define the mean range $\bar\ell_i$ for l-bit $i$, from the weighted (by the norm of the operator) average of the range of all of its
constituent p-bit product operators.  And we can define the average range for a given choice of l-bit operators as the average of the range over all the l-bits.
%\tred{Another measure of localization of l-bits can be made based on how well the Pauli "sum rule" $Tr [\tau^a_j]^2=1$ is saturated should the trace be carried over all p-bit product operators in the finite (small) neighborhood of $j$ of size $r$, i.e. $1-Tr [\tau^a_j]_r \equiv P_{aj}(r)\approx exp(-\kappa |r|)$.}
Of course, other definitions of the average range that are different in their details can be formulated and might be more practical and/or appropriate under some circumstances.

We expect that for a generic Hamiltonian in the many-body localized phase, if all many-body eigenstates of $H$ are localized there do exist one-to-one assignments from the eigenstates of $H$ to the eigenstates of the l-bits that give
a finite average range in the thermodynamic limit.
We want to choose the assignment that gives the minimum of the average range over all assignments, and this minimum will be a measure of the
localization length.  We expect that if we use this optimal assignment, the typical l-bit will consist of an infinite sum of p-bit product operators, but that the terms with long range will
have a total weight that typically falls off exponentially with the range.  Also there will be rare l-bits that have long mean range, due to rare ``resonances'', but these will occur with a probability that falls off exponentially with the range.  Again, the precise set-up of optimizing the ``range'' that we propose here can certainly be modified to something that is different in many details as long as it produces a useful definition of the l-bits and makes them well-localized.

For a generic ``nonintegrable'' short-range Hamiltonian outside of the many-body localized phase, it is expected that the eigenstates obey the Eigenstate Thermalization Hypothesis (ETH) \cite{eth1,eth2,eth3}.  For such systems, we expect that all $(2^N)!$ possible definitions of the l-bits will produce average ranges of order the size of the system, and that there are no operators that both commute with $H$ and are ``local'' in any sense of that word (other than $H$ itself and perhaps some finite number of other conserved quantities such as total spin or particle number).  Thus such Hamiltonians are not integrable in any useful or appropriate sense of the word ``integrable'' \cite{lych}, even though we can formulate a definition of an extensive set of (nonlocal) conserved quantities.

Our Hamiltonian $H$ is by assumption short-range when written in terms of the p-bits.  We also would like to discuss what $H$ looks like when written in terms of the l-bits.  Just as we can expand the l-bits in terms of the p-bit product operators, we can also expand any short-range p-bit operator that appears in $H$ in terms of the $4^N$ linearly-independent l-bit product operators.  If we are in the many-body localized phase, all many-body eigenstates of $H$ are localized, and we use an optimal definition of the l-bits, then we expect that the l-bit product operators that appear in $H$ will dominantly be of short range, with long-range l-bit operators typically having coupling constants in $H$ that vanish exponentially with the range. Furthermore, since $H$ commutes with all the $\tau^z_i$'s, no terms involving $\tau^x_i$'s or $\tau^y_i$'s may appear in $H$.  Thus the Hamiltonian is of the form
\begin{equation}
H=\sum_ih_i\tau^z_i + \sum_{i,j}J_{ij}\tau^z_i\tau^z_j + \sum_{i,j,k}K_{ijk}\tau^z_i\tau^z_j\tau^z_k +...
\end{equation}
when written in terms of the l-bit operators, with the couplings of the higher-order and longer-range terms in this expansion typically falling off
exponentially with the order and the range.

Traditional, translationally-invariant integrable one-dimensional models of $N$ spins have $N$ conserved local densities.  It appears that if you try to make other conserved quantities as composites (operator products) of these basic conserved densities, these are necessarily nonlocal operators of range $\sim N$.  For a many-body localized system, on the other hand, if we consider $n$ l-bits near site $i$, out of products of these l-bits, we can make $2^n$ independent conserved quantities that are localized near $i$.  In this sense, many-body localized systems have many more conservation laws that can affect local observables than do traditional translationally-invariant integrable systems.

Now let's consider the dynamics within the many-body localized phase. Sharp manifestations of localization are traditionally discussed in terms of the vanishing of DC transport.
% conductivity, i.e. absense of  low frequency macroscopic Kubo transport averaged systematically over some "diagonal" ensemble of eigenstates, e.g. Gibbs weight.
Much of the recent effort has focused instead on transient real time dynamics, often from a particularly simple initial product state of the bare degrees of freedom.
%with a definite fully-localized configuration of particle positions, spin etc. at $t=0$.
It follows from our general discussion above that such initial states of zero p-bit entanglement are %in fact
``wavepackets'' when expressed in terms of the many-body localized eigenstates of $H$, generally spread over a range of energies of order $\sim \sqrt{N}$.
Information about the real time dynamics of this initial state can be gleaned from the Hamiltonian written in terms of l-bits.  Most importantly, the presence of
%non-linearities (interactions)
interactions between the l-bits %Hamiltonian implies that the many-body spectrum cannot be written in terms of single particle levels and
means that generic initial states of the kind considered will dephase, so there will be no local observables that show long-time persistent oscillations.
%, i.e. even local few-body autocorrelations will not generically exhibit persistent oscillations, as would be expected, e.g. for the simple Anderson insulator
%whose dynamics is governed by the noninteracting l-bit Hamiltonian.  The many-body version of t
This dephasing occurs due to interactions generating entanglement between the interacting l-bits.

The dynamics of the l-bits in the many-body localized phase is in some sense simple: their $z$ components are frozen, while their transverse $xy$ components precess about the $z$ axes of their Bloch spheres at a rate that is set by the interactions with the $z$ components of all of the other l-bits. Thus they are subject to dephasing and decoherence due to the interaction with this static spin bath, but all $\tau^z_i$'s are conserved, so there is no ``dissipation''. Since the spin bath is static, the dephasing can in principle be reversed by spin echo procedures.  However, this requires having access to a single l-bit to apply the needed pulses to it without disturbing the other l-bits that are acting as the bath.  It is an interesting question to explore as to how strong a spin echo signal one could detect by instead doing the echo procedure on a ``bare'' p-bit, instead of the l-bit operator whose precise specification is nontrivial to find.

Next let's consider the spreading of entanglement within the many-body localized phase.  As in Refs. \cite{znid,bard,vosk}, start with an initial state at time $t=0$ that is a pure product state of the p-bits.  Thus it initially has no entanglement between p-bits.  This state is a very particular linear combination of the eigenstates of the Hamiltonian (and of the l-bits).  The eigenstates of the Hamiltonian each have short-range ``area law'' entanglement between the p-bits, while they are pure product states of the l-bits.  On a microscopic time scale, this initial linear combination of the eigenstates of $H$ will dephase, producing an area-law entanglement between the p-bits with a magnitude set by the typical entanglement in an eigenstate of $H$, as was seen and understood in the early time regime in Refs. \cite{bard,vosk}.

After this early-time transient, we can discuss what happens at later times in terms of the l-bits.  Each l-bit is precessing about its $z$-axis at a rate that is set by all the other l-bits that it interacts with.  Thus if its interaction with another l-bit is of magnitude $J$, the dependence of its phase on the state of that other l-bit will become significant once $Jt$ is of order one ($\hbar=1$).  Thus these two l-bits will become entangled after a time $\sim 1/J$.  Since the l-bit interactions $J$ in the localized phase fall off exponentially with distance, the number of other l-bits that a given l-bit is entangled with grows as $\sim \log^d{t}$ for a $d$-dimensional system.  The p-bits are composed of local l-bits, so their entanglement will also grow this way at long time, as seen in Refs. \cite{znid,bard}.  Thus although the localized many-body eigenstates of $H$ have only area-law p-bit entanglement, the dynamics of $H$ cause this logarithmic-in-time growth of the entanglement, which can continue without limit in an infinite system, due to the weak long-range interactions between the l-bits.  Note that in Ref. \cite{vosk}, they considered a special model that is at a random-singlet-type critical point within the many-body localized phase \cite{z2}, where the interactions instead fall off with distance as a ``stretched exponential'', allowing the entanglement to grow as a larger power of $\log t$.

To be slightly more quantitative about this spreading of entanglement, let's consider a generic many-body localized spin chain (not at a special critical point as in Ref. \cite{vosk}).  The localization length is $\xi$, so the typical interactions between l-bits fall off with distance $x$ as $J_{eff}(x) \approx J_0\exp{(-x/\xi)}$, where $J_0$ is an interaction scale at the lattice spacing.  The initial product states dephase in the limit of infinite times to produce entanglement entropy $s_{\infty}$ per spin for a finite block of consecutive spins.  If we then consider the long-time growth of the bipartite entanglement entropy between two semi-infinite half-chains, the distance the entanglement spreads is set by $J_{eff}(x) \sim 1/t$, or $x \sim \xi\log{(J_0t)}$.  The resulting entanglement entropy thus grows with time as $S \sim s_{\infty}\xi\log{(J_0t)}$.  This scenario seems quite consistent with the results reported in Ref. \cite{bard}.  Note that the value of $s_{\infty}$ will depend on the choice of initial states.  In Ref. \cite{bard} they chose initial states with the p-bits randomly oriented along their $z$ axes, which produces a rather small $s_{\infty}$, allowing the DMRG calculation to access fairly long times.  A much larger $s_{\infty}$ in the same model could be produced by instead orienting the p-bits initially perpendicular to their $z$ axes.

For some models that are less strongly disordered, we expect that there is a many-body mobility edge within the many-body spectrum of $H$, as is discussed in Ref. \cite{baa}.  In this event, almost all of the many-body eigenstates of $H$ are thermal and correspond to high temperatures (possibly including large negative temperatures), but we expect that there remain many-body localized eigenstates at the energies that would correspond to low temperatures.  Here the above construction defining l-bits clearly will not work, since it uses all of the eigenstates of $H$ and can work only if they are all localized.  But it does seem possible that some similar definition of l-bits should exist, somehow supplemented with restrictions on the local energy density in the vicinity of that l-bit so its localization length remains finite.  The difficulty will be to properly deal with the rare regions where the local energy density approaches that of the mobility edge (a new type of ``Griffiths singularity'').  We leave this challenge for future work.

We are grateful to A. Pal, T. Spencer, J. Imbrie, E. Altman, I. Cirac, J. Moore, F. Essler, G. Refael, S. Sondhi, M. Mueller, R. Vosk, R. Nandkishore, B. Altshuler, I. Aleiner and D. Basko for enlightening discussions.  This work was supported in part by NSF under DMR-0819860 (DAH) and DMR-0955714 (VO), and by funds from the DARPA optical lattice emulator program (DAH).

\end{document}